# "Tsessevich" Project: an Attempt to Find the System YY Dra. I

NATALIA A. VIRNINA

Department "High and Applied Mathematics"
Odessa National Maritime University, Odessa, Ukraine, virnina@gmail.com

**Abstract:** We present the first results of the project "Tsessevich". The main goal of this project is to find the real coordinates of eclipsing variable star YY Dra, which had been discovered by Tsessevich in 1934. However, the possible misprint in the coordinates entailed that the star had been lost and later identified with a cataclysmic star DO Dra, which located closely to the published coordinates of YY Dra.

We expected, that search for YY Dra will give us some new variable stars as a "by product". We plan to publish the results of the search as soon as the new variable stars would be confirmed. As the present paper is the first of the series of the papers about the results of this project, here we describe it's idea, used instruments, methods and computer programs. Also we present the first five new eclipsing W UMa-type variable stars. They are already registered in the VSX catalog. For these systems we've determined all parameters needed for the General Catalog of Variable Stars.

## 1. Introduction

In 1934, using the Moscow plates, Vladimir P. Tsessevich discovered an Algol-type variable star, which got the preliminary number SVS 504 Dra. He published it in the journal „Peremennye Zvezdy" with other 3 new variable stars, where he indicated that there is a new Algol-type variable star with the coordinates $\alpha_{1855} = 11^h35^m15^s$, $\delta_{1855} = +72°30'.4$ with the ephemeris $T_{min}=2419852.4+4^d.21123 \cdot E$ and the brightness range $12.9^m$ - $<14.5^m$ (Tsessevich, 1934). Later in the General Catalog of Variable Stars this star had been named YY Dra.

The object with the coordinates mentioned above, nowadays should have the coordinates $\alpha_{2000} = 11^h43^m34.41^s$, $\delta_{2000} = +71°42'9.4''$. But as there is no visible star on the spot with these coordinates, the suggestion about a misprint had been made. Unfortunately, no finding chart was published, which could allow to identify the field.

Andronov et al. (2008) presented a historical review. A story of YY Dra is not simple. Close to these coordinates, an X-ray source 2A 1150+720 was detected and later classified as a cataclysmic variable star by Patterson et al. (1982). It was also detected as a cataclysmic variable in the Palomar-Green Survey, and listed as PG 1140+719 (Green et al. 1982). They suggested an identification of this object with a previously registered variable YY Dra. Than Wenzel (1983) failed to find variability of the $12^m$ star close to the published position of YY Dra, but found an eruptive object at the position of PG 1140+719. He detected this object on only two plates from 700, and thus classified the object as a dwarf nova with an extremely long cycle length. This is definitely not an eclipsing variable with a well-defined period, and thus a separate official GCVS name "DO Dra" was assigned to PG 1140+719 (Kholopov et al. 1985; Samus' et al. 2007). The designation of the star was discussed by Patterson & Eisenman (1987). From a formal point of view, the situation is clear – there is a well known cataclysmic variable DO Dra, and a "lost" eclipsing variable which has a separate official GCVS name YY Dra. However, in the literature, DO Dra is sometimes still referred to as "YY Dra" or even "DO/YY Dra".

In the private communication with N. N. Samus about the true position of YY Dra, V. P. Tsessevich indicated the Moscow plates centered on another Algol-type variable star, Z Dra, and mentioned some "half of degree" position, which had been interpreted as the distance between Z Dra and YY Dra. Unfortunately, most of the plates of this region were gone during the Second World War.

Thus, as the true position of YY Dra, discovered by Tsessevich, is still a mystery, we've decided to search for this star. We named the research project "Tsessevich".

## 2. Instruments and fields

As the position of Tsessevich's star YY Dra on the plate is rather unsure, we decided to choose the survey telescope and several fields, which cover the whole plate 10°x10° centered on Z Dra. We used the remotely controlled refractor FSQ-106 (D=106mm, F=540mm) of Tzec Maun Observatory (New Mexico, USA). This telescope was combined with the CCD camera SBIG STL-11000M with an LRGBHα filter wheel. The field of view was 233.8′x155.9′, the image scale was 7.04"/pixel. Thus, we chose 12 fields, which slightly overlap each other and completely cover the rectangle 10°x10° centered on Z Dra ($\alpha_{2000}$ = $11^{h}45^{m}29.205^{s}$, $\delta_{2000}$ = +72°14′58.38″). The list of the centers of the fields with their numbers is presented in the Table 1. Also at the beginning of the project there were obtained a few series of images on the same telescope, centered right on Z Dra.

Table 1. List of centers of the fields.

| # | $\alpha_{2000}$ | $\delta_{2000}$ |
|---|---|---|
| 1 | $12^{h}42^{m}00^{s}$ | +76°00′00″ |
| 2 | $11^{h}45^{m}30^{s}$ | +76°00′00″ |
| 3 | $10^{h}49^{m}00^{s}$ | +76°00′00″ |
| 4 | $12^{h}34^{m}14^{s}$ | +73°30′00″ |
| 5 | $11^{h}45^{m}30^{s}$ | +73°30′00″ |
| 6 | $10^{h}56^{m}46^{s}$ | +73°30′00″ |
| 7 | $12^{h}27^{m}59^{s}$ | +71°00′00″ |
| 8 | $11^{h}45^{m}30^{s}$ | +71°00′00″ |
| 9 | $11^{h}03^{m}01^{s}$ | +71°00′00″ |
| 10 | $12^{h}23^{m}14^{s}$ | +68°30′00″ |
| 11 | $11^{h}45^{m}30^{s}$ | +68°30′00″ |
| 12 | $11^{h}07^{m}46^{s}$ | +68°30′00″ |

As there are no photometrical filters on this telescope, all images were obtained with a "clear" ("Luminance") filter. To determine the dependence between our photometrical system and the standard one, we've used the calibration of the stars in the vicinity of DO Dra, given by Henden (2010). 27 stars with different color indexes were involved, and we've found that our system is very close to the standard V-band. The dependence between our system and the standard V and R bands has been found as a linear polynomial using the program MCV (Andronov & Baklanov, 2004):

$$CV - V = -0.007(\pm 0.015) + 0.034(\pm 0.031) \cdot (V - R),$$

where $CV$ – our magnitudes, $R$ and $V$ – magnitudes in standard $R$- and $V$-bands. The values of the transformation coefficients are close to zero, thus we use the $V$ magnitudes for calibration of photometry.

## 3. Photometry

For calibration of the photometry, we've used Henden's (2007) data of the magnitudes of the stars in the vicinity of DO Dra. For the photometry of the new stars, studied in the present paper, we have chosen 3 stars: USNO-B1.0 1615-0090723, USNO-B1.0 1615-0091065 and USNO-B1.0 1617-0093328. Their coordinates and magnitudes are given in the Table 2 and their positions are marked on the Figure 1. However, only one field covers this region with the standard magnitudes. There are other two variable stars surrounded by known standard (Henden's) stars: the Mira-type star VX UMa ($\alpha_{2000}$ = $10^{h}55^{m}40.857^{s}$, $\delta_{2000}$ =

+71°52′10.06″) and SRb-type star SS Dra ($\alpha_{2000}$ = $12^h26^m21.031^s$, $\delta_{2000}$ = +68°41′16.64″). But as SS Dra is rather bright, only the faintest stars, given in the standards, could be used. For all other fields, we are going to measure the "secondary standard", which means that for other fields we'll determine the magnitudes of nearby comparison stars, using known standard magnitudes.

In further papers of this series, we'll indicate in each case for each field, which stars will be used as a "secondary standard".

Table 2. Reference stars

| # | USNO-B1.0 | $\alpha_{2000}$ | $\delta_{2000}$ | V mag | (V-R) mag |
|---|---|---|---|---|---|
| 1 | 1615-0090723 | $11^h41^m38.166^s$ | +71°35′54.95″ | 12.183 | 0.481 |
| 2 | 1615-0091065 | $11^h44^m27.327^s$ | +71°31′33.89″ | 13.390 | 0.350 |
| 3 | 1617-0093328 | $11^h42^m21.356^s$ | +71°43′48.86″ | 13.835 | 0.337 |

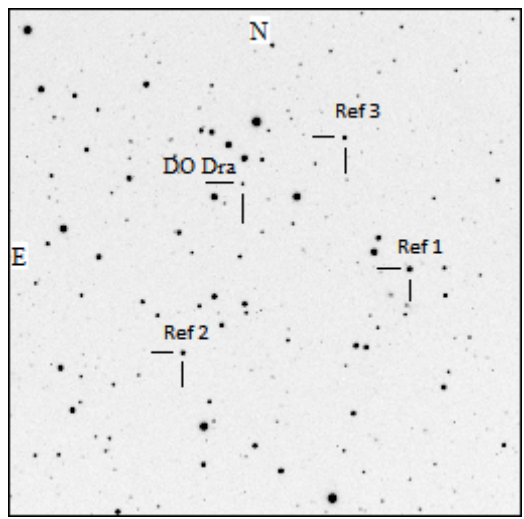

Fig. 1. 30′x30′ field of view. DO Dra and the reference stars are marked.

## 4. Data Analysis

To investigate the fields and to find the new variable stars, we've used the VaST software (developed by K. Sokolovsky and A. Lebedev, described by Kolesnikova et al., 2008) based on the SExtractor routine (Bertin & Arnouts, 1996). To identify possible variable stars among all stars of the image, we created a "r.m.s. scatter versus mean magnitude" diagram. The fainter objects tend to have larger scatter. Those objects, which bounce from this relation, (objects with a value of the r.m.s. scatter significantly larger than the typical value for their mean magnitude) are good candidates to be variable stars. False positives could also be caused if object is a galaxy or blended, either by the presence of a close companion or image defect etc.

We use the VaST software for each field at least twice: for the first time applying to the longest series (5 hours or more), which allowed us to discover new variables with fast variability (usually the short periodic variable stars); and for the second time we look for the slow variability and for the long periodic variables, using a few images from every observational set, obtained during the project.

To check if the found variables are still unknown, we use the VizieR searching database (Ochsenbein, Bauer & Marcout, 2000) of astronomical catalogues operated by the Centre de Donnees de Strasbourg.

For the preliminary determination of the period of new variable stars, we use the software "WinEffect" (Goransky, 2005). Then the software FDCN (Andronov, 1994, 2003) is used to determine and the coefficients of the statistically optimal trigonometric polynomials using the least squares method routine. This program yields the differential corrections for the period, initial epoch, and also the depth of minima and the height of maxima, and the asymmetry (for the pulsating stars), with the corresponding error estimates for each parameter. Thus, using this program for the periodical stars, for which phase curves are completely covered, we can determine all parameters needed for the General Catalog of Variable Stars (Samus' et al., 2009).

## 5. First results

In this paper we present five new variable stars, which have been found within the "Tsessevich" project. All these stars are short periodic eclipsing binary systems of W UMa-type. Among other new variables, they were discovered on the field, centered on Z Dra. Their positions and the position of Z Dra are marked on the chart, which covers the whole field of view and attached to this paper as Appendix 1. All presented stars we preliminary registered in the Variable Stars Index (VSX) catalog operated by AAVSO. The list of the coordinates of the new variable stars, their USNO-B1.0, 2MASS and VSX names are given in the Table 3.

The phase curves are shown on the Figures 2 – 6 respectively. All photometrical parameters of the new binary systems are summarized in the Table 4 with corresponding errors estimates.

All photometry data are given in the Appendix 2.

### 5.1 VSX J120641.2+713246

Using the FDCN software, we've determined the degree of statistically optimal trigonometric polynomial fit for the star VSX J120641.2+713246: $s = 8$. We've noted the presence of the total eclipsing in the secondary minimum. An approximate duration of the total eclipse is 0.112 of phase, which corresponds to 0.053242 d or 77 minutes.

### 5.2 VSX J114030.0+711102

For the star VSX J114030.0+711102, the degree of the statistically optimal trigonometric polynomial fit is $s = 6$. The depths of minima are nearly equal, but the difference is still noticeable, this gave us a possibility to determine which of minima is a primary one. One could notice a slight asymmetry of the shape of the secondary minimum, which could be caused by a spot.

### 5.3 VSX J115558.2+730025

The smoothing trigonometric polynomial fit of the degree $s = 8$ had been obtained for VSX J115558.2+730025. Both minima are rather deep and narrow, which phenomenologically kinships this system with Algol-type systems. However, the period of this star is short, $0.29404^d$, the shape of maxima is smoothed and it is rather complicated to determine the moments of the beginning and the end of the eclipses, which more suits to the classical description of EW-type stars. Thus, we classified this star as EW-type system.

### 5.4 VSX J114836.5+710751

For this star, the trigonometric polynomial fit of the order $s=8$ is the most optimal too. The photometrical characteristics of this star correspond to that ones of classical EW-type stars.

### 5.5 VSX J113727.2+722403

This star has nearly sinusoidal shape of the curve, except the difference between the magnitudes in maxima, thus the degree of the statistically optimal polynomial fit was $s=4$. In the first maximum the magnitude is 13.879±0.003 mag, in the second 13.848±0.003 mag, and this last value is given in the Table 4 as the maximum magnitude. The difference between the heights of maxima is 0.031±0.004, i.e. 7.75σ, with a corresponding "false alarm probability" of $1.1 \cdot 10^{-4}$, thus this phenomenon is statistically significant, so we conclude that the O'Connel effect is present.

Table 3. Cross-identification for new variable stars

| # | $\alpha_{2000}$ | $\delta_{2000}$ | USNO-B1.0 | 2MASS | VSX |
|---|---|---|---|---|---|
| 1 | $12^h06^m41.287^s$ | +71°32′46.93″ | 1615-0092898 | 12064133+7132468 | J120641.2+713246 |
| 2 | $11^h40^m30.022^s$ | +71°11′02.44″ | 1611-0091333 | 11403001+7111021 | J114030.0+711102 |
| 3 | $11^h55^m58.219^s$ | +73°00′25.33″ | 1630-0091892 | 11555825+7300249 | J115558.2+730025 |
| 4 | $11^h48^m36.524^s$ | +71°07′51.14″ | 1611-0091801 | 11483649+7107507 | J114836.5+710751 |
| 5 | $11^h37^m27.240^s$ | +72°24′03.43″ | 1624-0096365 | 11372726+7224033 | J113727.2+722403 |

Table 4. Parameters of the new binary systems. All magnitudes are given in V-band

| # | $\min_I$ | $\min_{II}$ | Max | Period, d | Initial epoch, HJD |
|---|---|---|---|---|---|
| 1 | 13.837±0.003 | 13.776±0.003 | 13.512±0.003 | 0.475371±0.000010 | 2455192.6379±0.0028 |
| 2 | 14.366±0.003 | 14.351±0.003 | 13.998±0.003 | 0.434911±0.000009 | 2455192.4773±0.0005 |
| 3 | 16.028±0.009 | 15.942±0.008 | 15.217±0.008 | 0.294040±0.000004 | 2455191.4051±0.0003 |
| 4 | 15.321±0.007 | 15.249±0.005 | 14.748±0.005 | 0.376832±0.000006 | 2455191.7896±0.0006 |
| 5 | 14.053±0.004 | 14.029±0.002 | 13.848±0.003 | 0.413660±0.000013 | 2455192.8152±0.0012 |

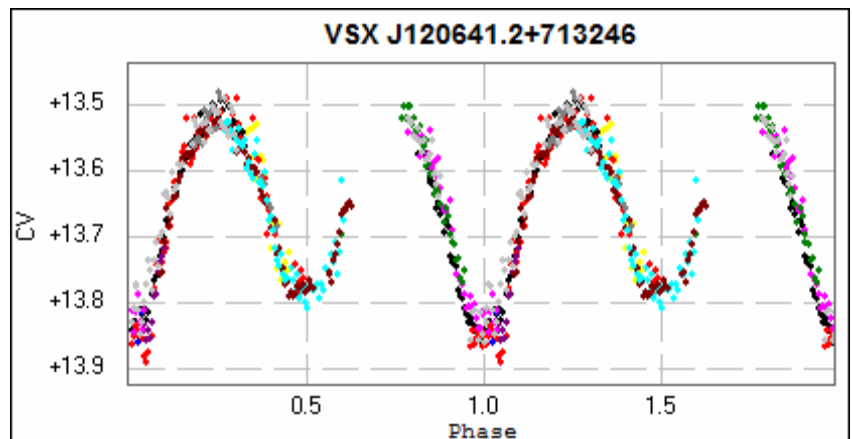

Fig. 2. Phase curve of the star USNO-B1.0 1615-0092898 = 2MASS 12064133+7132468 = VSX J120641.2+713246

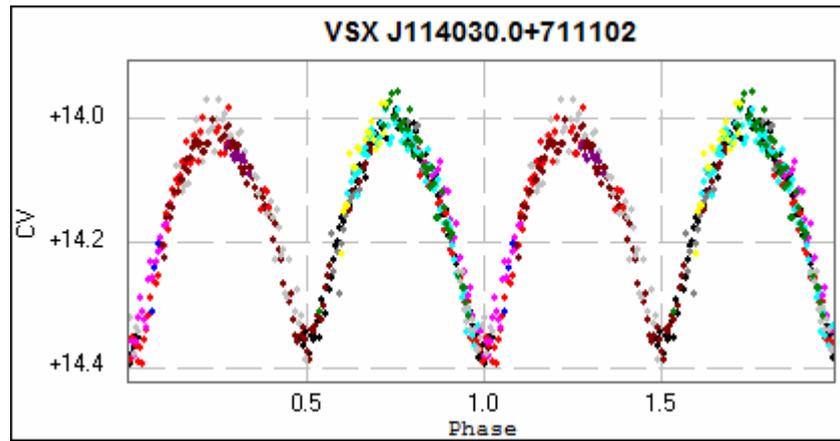

Fig. 3. Phase curve of the star USNO-B1.0 1611-0091333 = 2MASS 11403001+7111021 = VSX J114030.0+711102

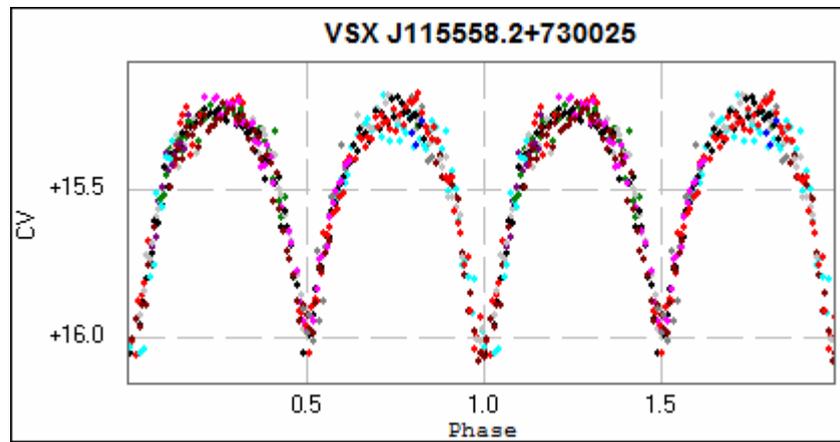

Fig. 4. Phase curve of the star USNO-B1.0 1630-0091892 = 2MASS 11555825+7300249 = VSX J115558.2+730025

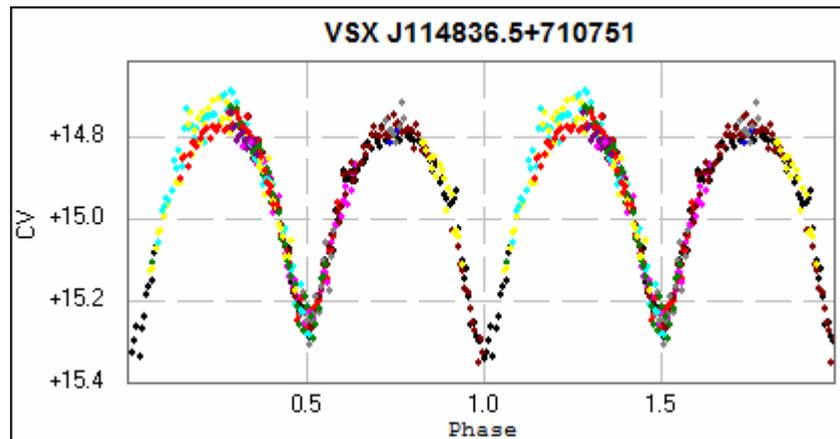

Fig. 5. Phase curve of the star USNO-B1.0 1611-0091801 = 2MASS 11483649+7107507 = VSX J114836.5+710751

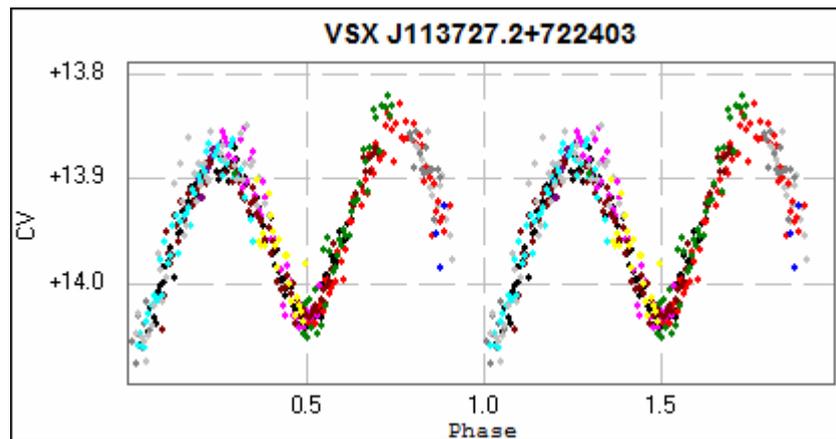

Fig. 6. Phase curve of the star USNO-B1.0 1624-0096365 = 2MASS 11372726+7224033 = VSX J113727.2+722403

## 6. Acknowledgments.

Author is thankful to Prof. Andronov I.L. for the idea of the project and for the helpful discussion.

This research is based on data collected with the Tzec Maun Observatory, operated by the Tzec Maun Foundation. Special thanks to Ron Wodaski (director of the observatory) and Donna Brown-Wodaski (director of the Tzec Maun Foundation).

Author is also thankful to Oleg Maliy for his help in preparing this paper.